\documentclass[11pt]{article}

\usepackage{amsmath,mathrsfs,amsbsy,amsthm}
\usepackage{cite}
\usepackage{graphicx,epsfig}
\usepackage{latexsym,amssymb}
\usepackage{epsf}
\usepackage{ifpdf,lineno}
\usepackage{color}
\usepackage{geometry}

\begin{document}
	
	\title{{\bf Stability Catalyzer for a Relativistic Non-Topological Soliton  Solution}}
	
\author{M. Mohammadi \\
	{\small \texttt{physmohammadi@pgu.ac.ir}}}
	\date{{\em{Physics Department, Persian Gulf University, Bushehr 75169, Iran.}}}
	\maketitle

	\maketitle
	
\begin{abstract}
 For a real nonlinear Klein-Gordon  Lagrangian density with a special   solitary wave solution (SSWS), which is essentially unstable, it is shown how adding a proper additional massless  term could guarantee the energetically  stability of the SSWS, without  changing its dominant dynamical equation and other properties. In other words,  it is a stability catalyzer. The additional  term contains a parameter $B$, which brings about more stability for the SSWS at larger values. Hence, if one considers $B$ to be an extremely  large value, then any  other  solution   which is not very close to the free far apart SSWSs and the trivial vacuum state, require an infinite amount of energy to be created. In other words,  the possible non-trivial stable configurations of the fields with the finite total energies are any  number of the far apart SSWSs, similar to any number of  identical particles.
\end{abstract}


\section{Introduction}\label{sec1}

For decades,  the classical relativistic  field equations with  stable solitary wave solutions (solions\footnote{According to some well-known references such as  \cite{rajarama}, a solitary wave solution is a soliton when it reappears  without any distortion after collisions. The  stability is essentially    a necessary condition for a solitary wave solution to be a soliton. However, in this paper,  we only adopt  the stability condition for the definition of a soliton solution.}) have drawn the interest of many physicists   \cite{rajarama,Das,lamb,Drazin,TS}. In fact,  soliton solutions  behave like real particles as   they have the non-disperse  localized energy density functions and satisfy the standard relativistic energy-momentum relations. For example, the real nonlinear Klein-Gordon  (RNKG)\cite{mandel,peskin} systems in $1+1$ dimensions with kink (anti-kink) solutions \cite{rajarama,TS,phi45,OV,GH,MM1,MR,JRM1,JRM2,DSG1,DSG2,DSG3,MM2,waz,ana1,
ana2,Kink1,Kink2,Kink3,Kink4,Kink5,Kink6,Kink7,Kink8,Kink9,Kink10,Kink11,Kink12,Kink13,
Kink14,Kink15,Kink16,Kink17}, Skyrme model of baryons \cite{TS,SKrme,SKrme2,SKrme3,SKrme4} and 't Hooft Polyakov model which yields magnetic monopole solutions   \cite{rajarama,TS,toft,pol,MKP,TOF,CTOPO}
 in $3+1$ dimensions are three well-known  systems which yield stable solitary wave solutions or solitons. One should note that, all the three systems mentioned above, yield topological solitons and the topological feature  is the main reason  behind  their stability. With topological  solutions, there are generally  complicated conditions on the boundaries to have a multi particle-like solution. However,  with the non-topological solitary wave solutions,  each arbitrary multi particle-like  solution can be obtained easily  just by adding distinct  far apart  solitary wave solutions together.

There have  been many works on the non-topological solitary  wave solutions  so far  \cite{rajarama,Das,lamb,Drazin,TS,N1,N2,N3,N4,N5,N6,N7,N8,N9,N10, N11,N12,N13,Vak3,Vak4,Vak5,Vak6,Lee3,Scoleman,
R1,R2,R3,R4,R5,R6,R7,R8,VD}.  However, the famous relativistic non-topological solitary  wave solutions are   Q-balls \cite{Vak3,Vak4,Vak5,Vak6,Lee3,Scoleman,R1,R2,R3,R4,R5,R6,R7,R8,VD,JOP,Non,JOP2}.
With  non-topological solitary wave solutions, an  important   criterion for  examining   the stability  is  the classical (or  Vakhitov-Kolokolov) criterion \cite{Vak3,Vak4,Vak5,Vak6,Vak1,Vak2,Vak777}.  The classical stability criterion is based on  examining dynamical equations  when they are  linearized for  small fluctuations above the background of the solitary wave solution. If the linearized equation does not lead to any unstable growing mode, the solitary wave solution  is called  a stable solution classically.     There is another stability criterion  called the energetical stability criterion.   In fact,  a special  solitary wave solution is  energetically stable if any arbitrary variation above its background  leads to an increase in the total energy. In other words, an energetically stable solitary wave solution would be stable against any arbitrary deformation \cite{Derrick,MM4,MMM1,MM5,MMM2}.

A solitary wave solution which is energetically stable would be a single solution among the  other (close) solutions. For example, the   kinks (anti-kinks) as the well-known topological solitary wave solutions  of the real nonlinear Klein-Gordon (RNKG) systems are inevitably energetically stable \cite{rajarama,Derrick}. However, a solitary wave solution, which is classically  stable, is not necessarily an energetically stable solution; or it is not a  single solution with the minimum  energy among the other (close) solutions. For example, some of the Q-ball solutions are classically stable \cite{Vak3,Vak4,Vak5,Vak6,Vak777}, but none of the them are energetically stable \cite{MM5}.  In fact, the energetical  stability criterion  is at a higher level  than the classical stability criterion. In other words,  if a solitary wave solution is  energetically  stable, it would undoubtedly be classically  stable, as well. Moreover, if a solitary wave solution is not  classically stable, it would not be  an energetically stable solution, as well.

In this paper, in line with the previous works \cite{MM4,MMM1,MM5,MMM2},  we   introduce a special  RNKG  model in $1+1$ dimensions with a well-formed non-topological  solitary wave solution which is essentially unstable  \cite{rajarama,Vak3}. But we will show  how adding a proper term  to the original  RNKG Lagrangian density, transforms the special solitary wave solution (SSWS) into  an energetically  stable  object. We call this additional term the “\textit{ stability catalyzer}", because  it behaves as  a massless spook\footnote{We chose  the word “\emph{spook}" so that it will not be confused with words like   “\emph{ghost}" and “\emph{phantom}", which have meaning in the  literature.}  which surrounds the SSWS  and guarantees its energetical stability. In other words,  it prevents any change in the internal structure of  the SSWS, and  leaves  the dominant dynamical equations  and other properties  of the SSWS invariant. It should be noted that, we  consider the model in $1+1$ dimension just for the sake of  simplicity, as it can be extended to $3+1$ dimensions with some modifications.

There is a  parameter $B$  in the stability  catalyzer term, which  leads to more stability for the SSWS at larger values.  In other words,  the larger the values the greater will be the increase in the total energy for any arbitrary small variation above the background of the SSWS.
Hence, if one considers   a system with an extremely  large  value of parameter $B$, then the other  configurations of the fields (which are not very  close to the SSWS and the vacuum state) need extremely  large energies  to be created; meaning that,
the possible solutions  of the  system with the finite energies are only  the  free  far apart SSWSs, as   multi   particle-like solutions.

The present  paper has been organized as follows: In the next section,     we  set up the basic equations for the RNKG systems with a single scalar field and consider a special RNKG model  with a special  non-topological non-vibrational  solitary wave solution which is essentially unstable. In section 3,  we  introduce the   stability catalyzer term, which serves   to introduce an  extended KG system\footnote{The extended KG systems were introduced in Ref.~\cite{MM4}. } with  a single non-topological energetically stable  SSWS. Section 4 presents an in-depth  study of the   stability of the SSWS for any arbitrary small deformations according to the energetical stability criterion.  The last section is devoted to  summary and conclusion.

\section{Single  field RNKG systems in $1+1$ dimensions}\label{sec2}

The simplest form of the real nonlinear Klein Gordon (RNKG) systems in $1+1$ dimensions  can be introduced by the following Lagrangian density:
\begin{equation} \label{lag}
{\cal L}_{o}=\partial^\mu \varphi\partial_\mu \varphi-U(\varphi),
\end{equation}
in which $\varphi$ is a single real scalar field  and $U(\varphi)$ is called  the field potential. Note that,  we set the speed of light to one ($c=1$) throughout the paper for the sake of simplicity. Using the principle of least  action,  the related equation of motion is
\begin{equation} \label{eom}
\Box \varphi=\frac{\partial^{2}\varphi}{\partial t^2}-\frac{\partial^{2}\varphi}{\partial x^2}=-\frac{1}{2}\frac{dU}{d\varphi}.
\end{equation}
Using the Noether's theorem \cite{mandel,peskin}, one can simply obtain the energy-momentum tensor:
\begin{eqnarray} \label{ed}
T^{\mu\nu}=2\partial^{\mu}\varphi\partial^{\nu}\varphi-g^{\mu\nu}{\cal L}_{o},
\end{eqnarray}
 in which $g^{\mu\nu}$  is the Minkowski metric. The $T^{00}$ ($T^{10}=T^{01}$) component of this tensor is the same energy  (momentum) density function, which for the Lagrangian density (\ref{lag}), is simplified   to
\begin{eqnarray} \label{emt}
T^{00}=\varepsilon_{o}(x,t)=\dot{\varphi}^2+\varphi'^2+U(\varphi)\quad (T^{01}=2\dot{\varphi}\varphi'),
\end{eqnarray}
in which the dot and the prime are symbols for time and  space derivatives respectively. The  integration of $T^{00}$ ($T^{10}$) over the whole space yields the same total energy $E$ (momentum $P$) of the system and always remains constant.

\begin{figure}[ht!]
   \centering
   \includegraphics[width=110mm]{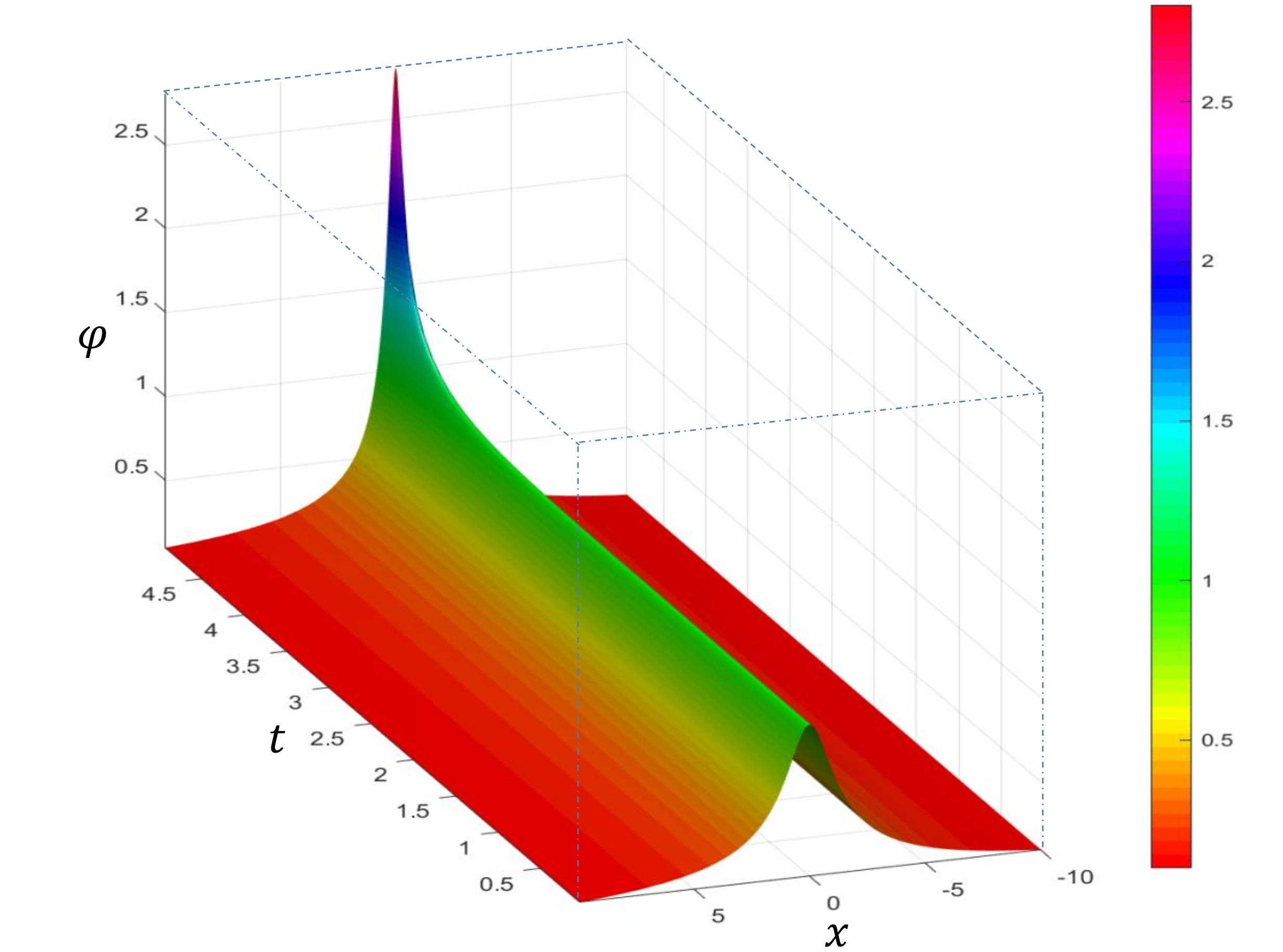}
   \caption{The non-topological static  solitary wave solution (\ref{ss}) of the RNKG system (\ref{po}) is essentially unstable and spontaneously blows  up. This figure shows a continuous representation of all the possible configurations of the spontaneous evolution of the  SSWS (\ref{ss}), which is initially considered to be at rest. It is obtained  from a finite difference method in Matlab for the  SSWS (\ref{ss}) as the initial condition  of the PDE (\ref{eom}).  } \label{pp}
\end{figure}

\begin{figure}[ht!]
   \centering
   \includegraphics[width=160mm]{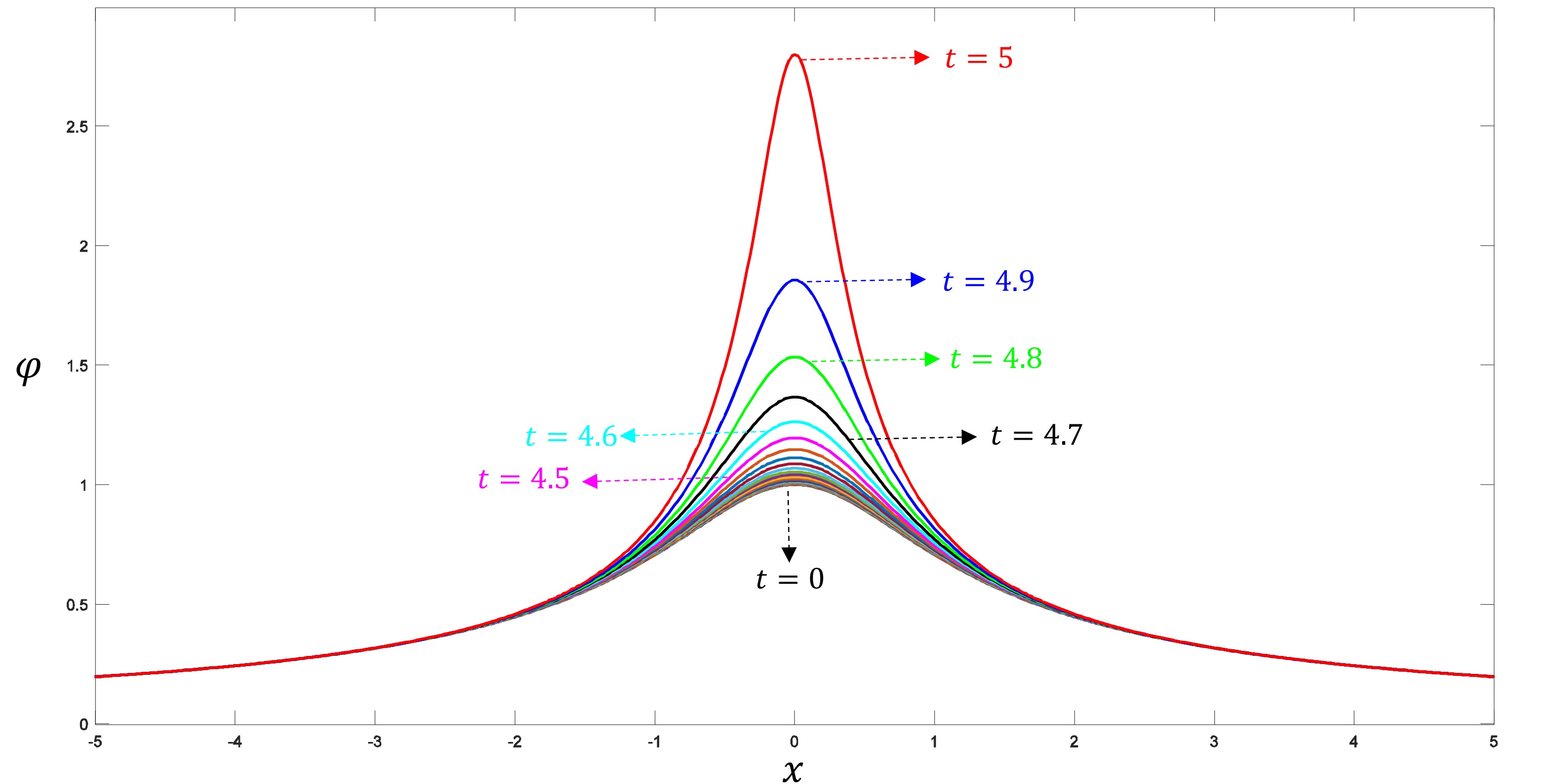}
   \caption{ Instead of Fig.~\ref{pp}, many configurations of the field at discrete times  in the range $0<t<5$  can be plotted in a 2D figure for better understanding. All  the different configurations have the same energy equaling the rest energy of the SSWS    (\ref{ss}) at $t=0$.} \label{3}
\end{figure}

In general, there is not a stable  non-topological non-vibrational solitary wave solution  for the RNKG systems in $1+1$ dimensions  \cite{rajarama,Vak3}. For example, if one considers a special  field potential  as follows:
\begin{equation} \label{po}
U(\varphi)= \varphi^4(1-\varphi^2),
\end{equation}
then the equation of motion (\ref{eom}), for a static (non-moving and  non-vibrational) solution $\varphi=\varphi_{o}(x)$,  is simplified   to
\begin{equation} \label{ste}
\frac{d^2 \varphi_{o}}{d x^2}=2\varphi_{o}^3-3\varphi_{o}^5,
\end{equation}
which has a  non-topological solution as follows:
\begin{equation} \label{ss}
\varphi_{o}(x)=\frac{\pm 1}{\sqrt{1+ x^2}}.
\end{equation}
Applying the Lorentz  transformations, the moving   version of this solution (\ref{ss}) can be obtained as well:
\begin{equation} \label{sss}
\varphi_{v}(x,t)=\varphi_{o}(\tilde{x})=\frac{\pm 1}{\sqrt{1+ \tilde{x}^2}},
\end{equation}
where $\tilde{x}=\gamma(x-vt)$, $\gamma=1/\sqrt{1-v^2}$ and $v$ is the velocity. However,  the field  potential (\ref{po}) for $|\varphi|>\frac{\sqrt{6}}{3}$ is  decreasing and for
$|\varphi|>1$ takes negative values. Therefore, the special solitary wave solution (\ref{sss}) is essentially unstable and without  violating  the conservation  energy law, the
effect of any small perturbation,  causes  the profile of the localized solution (\ref{ss}) to blow  up \cite{rajarama}(see Figs.~\ref{pp} and \ref{3}). In \cite{Vak3},  the instability of the non-topological non-vibrational solitary wave solutions  of the RNKG systems in $1+1$ dimensions  are generally  referred to the existence of the growing modes.

In general, since the theory is relativistic,  the same    well-known  relativistic relations between the moving and non-moving versions  of any arbitrary solution,  which has a  localized energy density function,   would be obtained,  meaning that:
 \begin{eqnarray} \label{d}
&&E=m = \int_{-\infty}^{+\infty} T^{00} dx=\int_{-\infty}^{+\infty} [\dot{\varphi_{v}}^{2}+\varphi_{v}'^2+U(\varphi_{v})]  dx=\gamma E_{o}=\gamma m_{o}, \\&&
P= \int_{-\infty}^{+\infty} T^{01}dx=2\int_{-\infty}^{+\infty} \dot{\varphi_{v}}\varphi_{v}' dx=\gamma m_{o} \textbf{v}.
\end{eqnarray}
where $E_{o}$ ($m_{o}$) is the same rest energy (mass) of the   solution. For the special solution (\ref{ss}), according to Eq.~(\ref{d}), the related  rest energy is $E_{o}=\pi/4$. Moreover, the width of any arbitrary moving solution is always smaller  than  its non-moving version, exactly according to the Lorentz contraction law.

\section{The stability catalyzer term for a SSWS}\label{sec3}

Here,  we attempt   to find a proper additional term for the original  Lagrangian density (\ref{lag}) that could  guarantee   the energetical  stability of the SSWS (\ref{ss}). However, similar to  a catalyzer,  we   expect that it has no role in the   dominant dynamical equation  and the other properties of the  SSWS (\ref{ss}). In other words, we first expect that the standard Eq.~(\ref{eom}) to remain  the dominant  dynamical equation  only  for the SSWS (\ref{ss}), and second,  the rest energy of the SSWS (\ref{ss}) to be a  minimum   among the energies of the  other (close) solutions. The  other complementary discussions are the same as those sufficiently presented in \cite{MM5}.

 However, we assume   a new extended KG  Lagrangian density  as follows:
\begin{equation} \label{lag2}
{\cal L}={\cal L}_{o}+F=\partial^\mu \varphi\partial_\mu \varphi-U(\varphi)+F,
\end{equation}
where $F$ is the same unknown additional (stability catalyzer) term, which should  be properly identified.
 We expect the new extended Lagrangian density  (\ref{lag2}) to be reduced  to the same original  version (\ref{lag}) just for the SSWS  (\ref{ss}), that is,   the new additional term $F$ should be zero only for the SSWS (\ref{ss}).  Note that,  the new extended system (\ref{lag2}) and the original RNKG system (\ref{lag}) are essentially different relativistic field  systems with   different solutions  except  for the SSWS (\ref{ss}), which is considered  to be a common solution. According to  the standard relativistic  Lagrangian densities  in physics, we expect  the unknown additional  scalar term $F$ to be  a function of the allowed scalars $\varphi$ and $\partial_{\mu}\varphi\partial^{\mu}\varphi$. However,  the new equation  of motion is
\begin{eqnarray} \label{new}
\left[\Box \varphi+\frac{1}{2}\frac{dU}{d\varphi}\right]+\frac{1}{2}\left[ \frac{\partial}{\partial x^{\mu}}\left(\frac{\partial F }{\partial (\partial_{\mu}\varphi)}\right)-\left(\frac{\partial F}{\partial \varphi}\right)  \right]=0.
\end{eqnarray}
For the SSWS (\ref{ss}) to  still remain  a solution of the new equation of motion (\ref{new}) (or  the new  equation of motion (\ref{new}) is reduced  to the same original version (\ref{eom})),  since the first part of this new equation, according to the  same original Eq.~(\ref{eom}), is satisfied automatically, i.e. $\left[\Box \varphi_{o}+\frac{1}{2}\frac{dU}{d\varphi_{o}}\right]=0$, and the functional $F$ is not essentially linear in  ${\cal L}_{o}$, so we conclude that the two distinct terms $\frac{\partial}{\partial x^{\mu}}\left(\frac{\partial F }{\partial (\partial_{\mu}\varphi)}\right)$ and $\frac{\partial F}{\partial \varphi}$ must be  zero independently for the SSWS (\ref{ss}).

To meet all these requirements, one can conclude that  $F$ must be a function of the powers of ${\cal K}$  (${\cal K}^{n}$'s with $n\geq3$), where    ${\cal K}$  is a  special scalar functional
\begin{equation}\label{k}
{\cal K}=\partial_{\mu}\varphi\partial^{\mu}\varphi+U(\varphi)=\dot{\varphi}^2-\varphi'^2+\varphi^{4}(1-\varphi^{2}),
\end{equation}
which is defined  to be zero  when we have a  SSWS (\ref{ss}).  For example, a simple choice   for the functional $F$ is
\begin{equation} \label{FF}
F=B{\cal K}^3,
\end{equation}
where $B$ is  a real positive number. For this special choice  (\ref{FF}), we obtain
\begin{eqnarray} \label{DF}
  && \frac{\partial}{\partial x^{\mu}}\left(\frac{\partial F }{\partial (\partial_{\mu}\varphi)}\right)= B{\cal K}\left[6{\cal K}\partial_{\mu}\partial^{\mu}\varphi+12\partial_{\mu}{\cal K}\partial^{\mu}\varphi \right], \nonumber  \\&&
\frac{\partial F}{\partial \varphi}=\left[3 B{\cal K}^{2}\frac{\partial {\cal K}}{\partial \varphi}\right],\nonumber
\end{eqnarray}
which are both obviously zero for the SSWS (\ref{ss}). In fact, each term on the right hand side of the above relations   contains a power of ${\cal K}$ and hence all are zero for the SSWS (\ref{ss}). Therefore, with this special choice (\ref{FF}), we can be certain  that the previous SSWS (\ref{ss}) is again a solution of the new extended system (\ref{lag2}), and the new dynamical equation (\ref{new}) is reduced to the same original  one (\ref{eom}), as its dominant dynamical equation.

The  energy density functional  of the new extended system (\ref{lag2}) can be obtained  easily
\begin{eqnarray} \label{nte}
T^{00}=\frac{\partial{\cal L}}{\partial \dot{\varphi}}\dot{\varphi}-{\cal L}=\left[\dot{\varphi}^{2}+\varphi'^2+U(\varphi)\right]+\left[B{\cal K}^{2}(6\dot{\varphi}^{2}-{\cal K})\right]=\varepsilon_{o}+\varepsilon_{1}.
\end{eqnarray}
According to Eq.~(\ref{k}), the second part of the energy density  (\ref{nte}) becomes
\begin{equation} \label{e1}
\varepsilon_{1}=B{\cal K}^{2}(5\dot{\varphi}^{2}+\varphi'^2+\varphi^{4}(\varphi^{2}-1)),
\end{equation}
which is zero  for the SSWS (\ref{ss}) and the vacuum state $\varphi=0$. However, it is not   a positive definite function,  because function $\varphi^{4}(\varphi^{2}-1)$ in the range $0<|\varphi|<1$ is negative. Hence, we cannot be certain   about the energetical  stability of the  SSWS (\ref{ss}).

In order to introduce  a new proper  additional term $F$ for which  the energetical stability of the SSWS (\ref{ss}) is properly guaranteed,  we have to use  a new  scalar field $\theta$ which  can be called the  phase field or the catalyzer field. However,  the new  proper additional  term $F$ can be introduced as follows:
\begin{equation} \label{FFF}
F=B\sum_{i=1}^{3}{\cal K}_{i}^3,
\end{equation}
where
 \begin{eqnarray} \label{e5}
  &&{\cal K}_{1}= \varphi^4 \mathbb{S}_{1},\\ \label{e52}&&
{\cal K}_{2}= \varphi^4 \mathbb{S}_{1}+\mathbb{S}_{2}, \\ \label{e53}&&
{\cal K}_{3}= \varphi^4 \mathbb{S}_{1}+\mathbb{S}_{2}+2 \varphi^2 \mathbb{S}_{3},
\end{eqnarray}
and
\begin{eqnarray} \label{sc2}
  &&\mathbb{S}_{1}=\partial_{\mu}\theta\partial^{\mu}\theta-1,\\ \label{sc22}&&
 \mathbb{S}_{2}=\partial_{\mu}\varphi\partial^{\mu}\varphi+\varphi^{4}(1-\varphi^{2}),\\ \label{sc23}&&
 \mathbb{S}_{3}=\partial_{\mu}\varphi\partial^{\mu}\theta.
\end{eqnarray}
In general, since  $\mathbb{S}_{i}$'s   are three independent scalars, it is not possible for them to be zero simultaneously except for  the non-trivial SSWS (\ref{ss}) with $\theta=\omega_{s}t= \pm t$. In other words, $\mathbb{S}_{1}=0$, $\mathbb{S}_{2}=0$ and $\mathbb{S}_{3}=0$ are three independent coupled nonlinear PDE's which do not have any non-trivial  common solutions except for the SSWS (\ref{ss}) with $\theta=\pm t$ (see the Appendix A). In fact, the same result applies to ${\cal K}_{i}$'s,  since ${\cal K}_{i}$'s are three independent linear combinations of the scalars  $\mathbb{S}_{i}$'s,  they are not zero simultaneously  except for  the SSWS (\ref{ss}) with  $\theta=\pm t$. Note that, for a moving version of the SSWS (\ref{sss}), which moves at the velocity of $v$, the  proper phase function $\theta$, for which all $\mathbb{S}_{i}$'s  would be zero simultaneously,  is $\theta=k_{\mu}x^{\mu}$, i.e. the boosted version of  function $\theta= \pm t$, provided
\begin{equation} \label{pro}
k^{\mu}\equiv (k^{0},k^{1})=(\omega,k)=(\omega,\omega v),
\end{equation}
where $\omega=\gamma\omega_{s}$ and $\omega_{s}=\pm 1$.

However, the dynamical equations of motion of the extended KG system  (\ref{lag2}) with the new additional term (\ref{FF}) can be obtained easily as follows:
\begin{eqnarray} \label{geq}
&&\Box \varphi+\dfrac{1}{2}\frac{dU}{d\varphi}+\frac{1}{2}\left[ \frac{\partial}{\partial x^{\mu}}\left(\frac{\partial F }{\partial (\partial_{\mu}\varphi)}\right)-\left(\frac{\partial F}{\partial \varphi}\right)  \right]=0,\\ \label{geq2} &&   \frac{\partial}{\partial x^{\mu}}\left(\frac{\partial F }{\partial (\partial_{\mu}\theta)}\right)=0.
\end{eqnarray}
Again, it is easy to show that all the different first and second derivatives of   $F$ (\ref{FFF}), which were seen in the Eqs.~(25) and (26),   for the SSWS (\ref{ss}) with $\theta=\pm t$, would be zero simultaneously. In other words, for the SSWS (\ref{ss}) with $\theta=\pm t$, Eq.~(26)  is automatically  satisfied  and  Eq.~(25)  is reduced to the same standard original version (\ref{eom}) as the dominant dynamical equation of the free SSWS (\ref{ss}). Note that the SSWS (\ref{ss}) in the new extended system (\ref{lag2}) must be now considered along with a    scalar field $\theta=\pm t$.   However, from here on in this paper,  the non-moving SSWS is as follows:
 \begin{equation} \label{SSWS}
\varphi_{s}(x)=\varphi_{o}(x)=\frac{\pm 1}{\sqrt{1+ x^2}},\quad\quad \theta_{s}(t)=\omega_{s}t=\pm t.
\end{equation}
Hence, the   moving version of the SSWS (\ref{SSWS}) would be
 \begin{equation} \label{SSWS2}
\varphi_{v}(x,t)=\varphi_{s}(\tilde{x})=\frac{\pm 1}{\sqrt{1+ \tilde{x}^2}},\quad\quad \theta_{v}(x,t)=\theta_{s}(\tilde{t})=k_{\mu}x^{\mu}=\omega t-kx.
\end{equation}
where $\tilde{t}=\gamma(t-vx)$ and $\tilde{x}=\gamma(x-vt)$.

The   energy density functional  of  the  extended  KG system  (\ref{lag2}) with the new additional  term (\ref{FFF})  is
\begin{eqnarray} \label{MTE}
&&\varepsilon(x,t)=T^{00}=\frac{\partial{\cal L}}{\partial \dot{\varphi}}\dot{\varphi}+\frac{\partial{\cal L}}{\partial \dot{\theta}}\dot{\theta}-{\cal L}=\varepsilon_{o}+\varepsilon_{1}+\varepsilon_{2}+\varepsilon_{3}=\nonumber\\&&
\quad\quad\quad\left[\dot{\varphi}^{2}+\varphi'^2+U(\varphi)\right]+B\sum_{i=1}^{3}\left[3C_{i}
{\cal K}_{i}^{2}-{\cal K}_{i}^3\right],
\end{eqnarray}
which are divided  into four distinct  parts and
\begin{equation}\label{cof}
C_{i}=\dfrac{\partial{\cal K}_{i}}{\partial \dot{\theta}}\dot{\theta}+\dfrac{\partial{\cal K}_{i}}{\partial \dot{\varphi}}\dot{\varphi}=
\begin{cases}
\quad\quad 2\varphi^4\dot{\theta}^{2} &  i=1
\\
2(\dot{\varphi}^{2}+\varphi^4\dot{\theta}^2) & i=2
\\
2(\dot{\varphi}+\varphi^2\dot{\theta})^{2}
 & i=3.
\end{cases}
 \end{equation}
 After a straightforward calculation, one can obtain:
 \begin{eqnarray} \label{eis}
&&\varepsilon_{1}=B{\cal K}_{1}^2[5\varphi^4\dot{\theta}^2+\varphi^4\theta'^2+\varphi^4]\geq 0,\\ \label{eis2}&&
\varepsilon_{2}=B{\cal K}_{2}^2[5\varphi^4\dot{\theta}^2+5\dot{\varphi}^2+\varphi^4\theta'^2+ \varphi'^2+\varphi^{6}]\geq 0, \\ \label{eis3}&&
\varepsilon_{3}=B{\cal K}_{3}^2[5(\varphi^2\dot{\theta}+\dot{\varphi})^2+(\varphi^2 \theta'+ \varphi')^2+\varphi^{6}]\geq 0.
\end{eqnarray}
All terms in the above relations are now positive definite, therefore all the $\varepsilon_{i}$'s ($i=1,2,3$) are positive definite functions and bounded from below by zero. All $\varepsilon_{i}$'s ($i=1,2,3$) are zero simultaneously just for the trivial vacuum state $\varphi=0$ and the non-trivial SSWS (\ref{SSWS}), just as we expected  from the catalyzer. Now, if parameter  $B$
is considered  to be a  large number, since  at least one of the ${\cal K}_{i}$'s  is a non-zero  function for any other solution,  then at least one of the  $\varepsilon_{i}$'s ($i=1,2,3$), which all contain parameter $B$, would be a  large positive function. It means that for  other solutions (except for the ones which are very close to the vacuum $\varphi=0$), the related  energies are always larger than the rest energy of the SSWS (\ref{SSWS}). Unlike $\varepsilon_{i}$'s ($i=1,2,3$), which are three absolute positive functions and are minimum for the SSWS (\ref{SSWS}), $\varepsilon_{o}$ is not  an absolute positive function and is not a minimum for the SSWS (\ref{SSWS}).
In the next section, we will show  the role of  $\varepsilon_{o}$ in the stability considerations, meaning that if we take an extended system (\ref{lag2}) with a large parameter $B$ (approximately $B>10^2$), it would be  completely ineffective.

\section{the energetically stability of the SSWS}

In general, a solitary wave solution (such as kink and anti-kink solutions of the RNKG  systems)  is  energetically stable if its rest energy  is  at a minimum   among the energies of the other (close) solutions. In other words, for an energetically stable solitary wave solution, any arbitrary deformation (variation)  above the background, leads to an increase in the total energy. In this section, we specifically  examine  the energetical  stability of  the SSWS  (\ref{SSWS}). In fact,   we are going to consider the variation of the total energy versus  any arbitrary small deformation above the background of the SSWS  (\ref{SSWS}), which is at rest. In general, any deformed version of the  SSWS (\ref{SSWS}) can be introduced  as follows:
\begin{equation} \label{so1}
\varphi(x,t)=\varphi_{s}(x)+\delta \varphi(x,t) \quad \textrm{and} \quad \theta(x,t)=\theta_{s}(t)+\delta \theta(x,t),
\end{equation}
where $\delta \varphi$ and $\delta \theta$  are considered as  arbitrary small  functions  of space-time.
Now, if we insert  (\ref{so1}) in   $\varepsilon_{o}(x,t)$ and  keep the terms
up to the second  order of small variation $\delta \varphi$, then it yields
\begin{eqnarray} \label{so3}
&&\varepsilon_{o}(x,t)=\varepsilon_{os}(x)+\delta\varepsilon_{o}(x,t)\approx\varepsilon_{os}(x)+\delta\varepsilon_{o1}(x,t)+
\delta\varepsilon_{o2}(x,t)=\left( \varphi_{s}'^2+U(\varphi_{s})\right)+\nonumber\\&&
\quad\quad 2\left(\varphi_{s}'  (\delta \varphi')+\frac{1}{2}\frac{dU(\varphi_{s})}{d\varphi_{s}}(\delta \varphi)\right)+\left((\delta\dot{\varphi})^2+ (\delta\varphi')^2 +\frac{1}{2}\frac{d^2U(\varphi_{s})}{d\varphi_{s}^2}(\delta \varphi)^2\right)
\end{eqnarray}
where $\varepsilon_{os}$, $\delta\varepsilon_{o1}$ and $\delta\varepsilon_{o2}$   are defined on the right hand side of the above equation, respectively.  $\varepsilon_{os}(x)=\left( \varphi_{s}'^2+U(\varphi_{s})\right)$ is the energy density function of the non-moving SSWS  (\ref{SSWS}).  $\delta\varepsilon_{o1}$ and $\delta\varepsilon_{o2}$ are  functionals  of the  first and second  order of the small  variation $\delta \varphi$, respectively.
Note that, for a non-moving SSWS (\ref{SSWS}), $\dot{\varphi_{s}}=0$, $\theta_{s}'=0$ and $\dot{\theta_{s}}=\omega_{s}=\pm 1$. It is obvious that $\delta\varepsilon_{o1}$, $\delta\varepsilon_{o2}$, and hence  $\delta\varepsilon_{o}$ are not necessarily the positive definite small  functionals.
Now, one can do a similar procedure  for the additional terms $\varepsilon_{i}$'s ($i=1,2,3$).  If  one inserts   a slightly deformed  SSWS (\ref{so1}) in  $\varepsilon_{i}$ ($i=1,2,3$), it yields
\begin{eqnarray} \label{so4}
&&\varepsilon_{i}(x,t)=\varepsilon_{is}+\delta\varepsilon_{i}=\delta\varepsilon_{i}=B[3(C_{is}+\delta C_{i})({\cal K}_{is}+\delta{\cal K}_{i})^{2}-({\cal K}_{is}+\delta{\cal K}_{i})^{3}]=\nonumber\\&&
B[3(C_{is}+\delta C_{i})(\delta{\cal K}_{i})^{2}-(\delta{\cal K}_{i})^{3}]\approx B[3C_{is}(\delta{\cal K}_{i})^{2}-(\delta{\cal K}_{i})^{3}]\approx B[3C_{is}(\delta{\cal K}_{i})^{2}]>0\nonumber\\&&
\end{eqnarray}
in which $\varepsilon_{is}=0$, ${\cal K}_{is}=0$ and $C_{is}=\omega_{s}^2\varphi_{s}^4$ are related   to the SSWS (\ref{SSWS}). According to Eq.~(\ref{so4}), since $C_{i}>0$ (\ref{cof}),
$\delta\varepsilon_{i}$'s  are   positive definite,  as is generally  expected  from   Eqs.~(\ref{eis}), (\ref{eis2}) and (\ref{eis3}).

According to Eqs.~(\ref{e5})-(\ref{sc23}),  keeping up the terms to  the first  order of small variations for the deformed SSWS (\ref{so1}),  we have
\begin{eqnarray} \label{sdd}
&&\delta{\cal K}_{1}\thickapprox 2\omega_{s}\varphi_{s}^4\delta \dot{\theta},\nonumber\\&&
\delta{\cal K}_{2}\thickapprox \delta{\cal K}_{1}-2\varphi_{s}' (\delta \varphi')-2(3\varphi_{s}^5-2\varphi_{s}^3)\delta \varphi,\nonumber\\&&
\delta{\cal K}_{3}\thickapprox \delta{\cal K}_{2}+2\varphi_{s}^2(\omega_{s}\delta\dot{\varphi}-\varphi_{s}'\delta \theta').
\end{eqnarray}
Since $\delta{\cal K}_{i}$'s are linear in the first  order of small variations  $\delta\varphi$, $\delta\theta$ and their derivatives ($\delta \dot{\theta}$, $\delta \varphi'$, $\delta\dot{\varphi}$ and $\delta \theta'$), thus,  according to Eq.~(\ref{so4}),  $\delta\varepsilon_{i}$'s are positive definite linear functions of the second order of small variations and their derivatives,  which   are all multiplied by $B$.

For any arbitrary small variations  $\delta\varphi$ and $\delta\theta$  above the background of a non-moving  SSWS (\ref{SSWS}), the variation of the total   energy  can be calculated  by the integration of $\delta\varepsilon$  over the whole space:
\begin{equation}\label{Eo}
\delta E=\int_{-\infty}^{+\infty} (\delta \varepsilon) ~dx=\int_{-\infty}^{+\infty} (\delta\varepsilon_{o}+\sum_{i=1}^{3}\delta\varepsilon_{i}) ~dx=\sum_{j=0}^{3}\delta \mathbb{E}_{j}.
\end{equation}
To show that the SSWS (\ref{SSWS}) is energetically  stable, we must  prove that $\delta E$ is always positive for any arbitrary small deformation. In other words, if any arbitrary deformation needs external energies to occur, then the SSWS (\ref{SSWS}) is an energetically stable solitary wave solution. Since $\delta\varepsilon_{1}$, $\delta\varepsilon_{2}$ and $\delta\varepsilon_{3}$ are positive definite  functions, then their  integration over the whole space, i.e. $\delta \mathbb{E}_{1}$, $\delta \mathbb{E}_{2}$ and $\delta \mathbb{E}_{3}$,  always leads  to positive  values. Now, let us to focus  on the $\delta \mathbb{E}_{o}$:
\begin{equation}\label{Eoo}
\delta \mathbb{E}_{o}=\delta \mathbb{E}_{o1}+\delta \mathbb{E}_{o2}=\int_{-\infty}^{+\infty}  \delta\varepsilon_{o1} ~dx+\int_{-\infty}^{+\infty} \delta\varepsilon_{o2}~ dx,
\end{equation}
where, $\delta \mathbb{E}_{o1}$   is the contribution of the first  order  of variations in  $\delta \mathbb{E}_{o}$, which  we will show that it would be zero in general.  For the not-deformed  non-moving SSWS (\ref{SSWS}), according to Eq.~(\ref{eom}),  as its  dominant  dynamical equation, we can use  $\varphi''_{s}=\frac{d^2\varphi_{s}}{dx^2}$ instead of $\frac{1}{2}\frac{dU(\varphi_{s})}{d\varphi_{s}}$  to obtain:
\begin{eqnarray} \label{sodf}
&&\delta\varepsilon_{o1}=
  2\left[ \varphi_{s}'  (\delta \varphi')+\frac{1}{2}\frac{dU(\varphi_{s})}{d\varphi_{s}}(\delta \varphi)\right]=  2\left[ \varphi_{s}'  (\delta \varphi')+(\delta  \varphi)\varphi''_{s}\right]=2\frac{d}{dx}\left(\delta\varphi\frac{d\varphi_{s}}{dx}\right).
\end{eqnarray}
Hence, the integration of $\delta\varepsilon_{o1}$ over the  whole space leads to
\begin{equation}\label{sxc}
\int_{-\infty}^{+\infty}\delta\varepsilon_{o1}~dx=\left.2(\delta\varphi\frac{d\varphi_{s}}{dx})\right|_{+\infty} -\left.2(\delta\varphi\frac{d\varphi_{s}}{dx})\right|_{-\infty}=0.
\end{equation}
Note that, $\delta\varphi$ and $\frac{d\varphi_{s}}{dx}$ are zero at $\pm\infty$.
Therefore, the following result is generally valid:
\begin{equation}\label{of}
\delta E=\int_{-\infty}^{+\infty}(\delta\varepsilon_{e}) ~dx=\int_{-\infty}^{+\infty} (\delta\varepsilon_{o2}+\sum_{i=1}^{3}\delta\varepsilon_{i})~dx,
\end{equation}
where $\delta\varepsilon_{e}=\delta\varepsilon-\delta\varepsilon_{o1}=\delta\varepsilon_{o2}+\sum_{i=1}^{3}\delta\varepsilon_{i}$, and can be called the effective variation of the energy density function.
Now, if one can prove that for all arbitrary small variations, $\delta\varepsilon_{e}$ always remains a positive function, then the integration of that would be  always positive as well, and the energetical stability condition is fulfilled.
Note that, $\sum_{i=1}^{3}\delta\varepsilon_{i}$ is essentially positive definite, but  $\delta\varepsilon_{o2}$ is not necessarily  a positive function.

Since $\frac{1}{2}\frac{d^2U(\varphi_{s})}{d\varphi_{s}^2}=-15\varphi_{s}^4+6\varphi_{s}^2>0$ for $|\varphi_{s}|<\frac{\sqrt{10}}{5}$, hence undoubtedly, $\delta\varepsilon_{o2}=(\delta\dot{\varphi})^2+ (\delta\varphi')^2 +\frac{1}{2}\frac{d^2U(\varphi_{s})}{d\varphi_{s}^2}(\delta \varphi)^2$ itself would be positive for the points $x$  that   $|\varphi_{s}(x)|$ is  less than  $\frac{\sqrt{10}}{5}$, and then $\delta\varepsilon_{e}>0$ for such points.
But,  function $\frac{1}{2}\frac{d^2U(\varphi_{s})}{d\varphi_{s}^2}$, for the points $x$ that    $|\varphi_{s}(x)|>\frac{\sqrt{10}}{5}$,  would be negative, and we cannot be certain  that $\delta\varepsilon_{o2}$ (and then  $\delta\varepsilon_{e}$) is always positive. Nevertheless, if one considers a system with a large value of parameter $B$,   we can be certain  that  $\delta\varepsilon_{e}>0$ for all points.  In fact, $|\delta\varepsilon_{o2}|$ is a function    of the second order of $\delta\varphi$, $\delta\varphi'$ and $\delta\dot{\varphi}$  which does  not contain parameter $B$,   but $\delta\varepsilon_{i}$'s ($i=1,2,3$) are also functions   of the second order of variations $\delta \varphi$, $\delta\theta$ and their derivatives  which are multiplied by $B$. Hence, we are certain   that always   $\sum_{i=1}^{3}\delta\varepsilon_{i}\gg |\delta\varepsilon_{o2}|$ or $\delta\varepsilon_{e}\approx \sum_{i=1}^{3}\delta\varepsilon_{i}>0$, provided that  $B$ is  a large number (approximately $B>10^2$). Accordingly, for the arbitrary variations $\delta\varphi$ and $\delta\theta$, $\delta E$ would be always positive and then we ensure that the SSWS (\ref{SSWS}) is  an energetically stable object, meaning that, its energy would be at a minimum among the  other (close) solutions. It should be noted  that, the   theory is relativistic, hence confirming the energetical stability of the SSWS at rest, is  generalized to all moving versions at any arbitrary speed.

\begin{figure}[ht!]
   \centering
   \includegraphics[width=160mm]{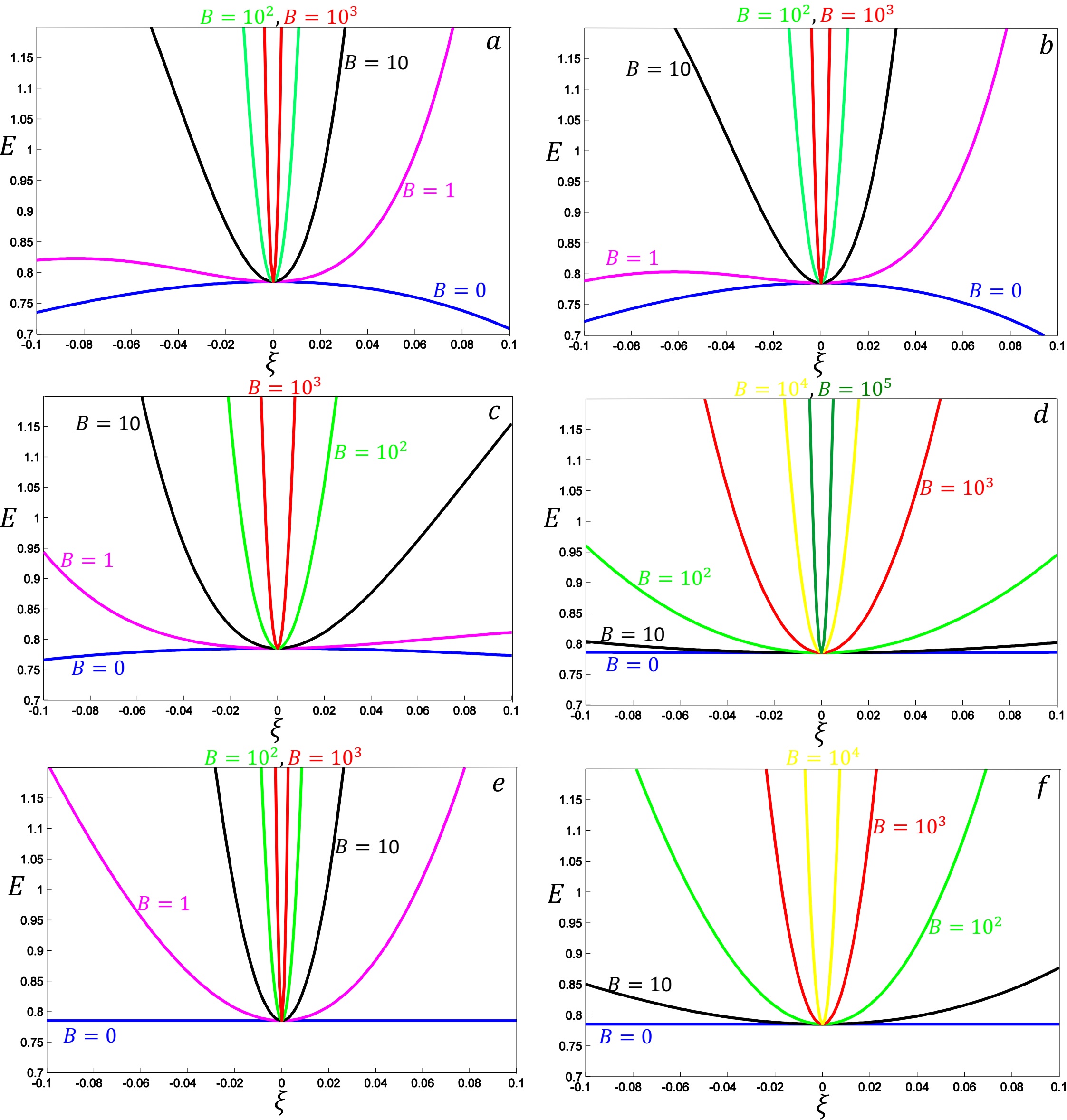}
   \caption{Variations  of  the total  energy $E$ versus small $\xi$  and different $B$'s at $t=0$ for the SSWS (\ref{SSWS}).  The Figs a-f are related to different variations (\ref{var1})-(\ref{var6}) respectively. Note that for the case $\xi=0$, in all figures,  the total energy is the same rest energy of the SSWS (\ref{SSWS}), i.e. $E(\xi=0)=E_{o}=\frac{\pi}{4}$. } \label{mmmm}
\end{figure}

To summarize, according to the previous considerations,   for any small deformation above the background of the SSWS (\ref{SSWS}), we finally have
\begin{equation}\label{bbn}
E=E_{s}+\delta E\approx\frac{\pi}{4}+\sum_{i=1}^{3}\int_{-\infty}^{+\infty}
\delta\varepsilon_{i} ~dx=\frac{\pi}{4}+3B\sum_{i=1}^{3}\int_{-\infty}^{+\infty}
[C_{is}(\delta{\cal K}_{i})^{2}]dx,
\end{equation}
where, $E$ is the total energy of the small deformed SSWS (\ref{so1}), and $E_{s}=\pi/4$ is the rest energy of the SSWS (\ref{SSWS}). It is true that for  the hypothetical  solutions (\ref{so1}), which are  close to the SSWS (\ref{SSWS}), the field variations $\delta\varphi$, $\delta\theta$ and hence  $\delta{\cal K}_{i}$'s  are  small, but the term  $[3B C_{is}(\delta{\cal K}_{i})^{2}]$ is not necessarily  small,  because it contains  the large parameter  $B$. Hence, $\delta E$ is not necessarily small as well (see Fig.~\ref{mmmm}).   For any  close solution (\ref{so1}), with two specific small variations  $\delta\varphi$ and $\delta\theta$, there are three  specific $\delta{\cal K}_{i}$'s that are used to obtain the total energy (\ref{bbn}). Since $\delta E$ is proportional to the integration of   $\sum_{i=1}^{3}C_{is}(\delta{\cal K}_{i})^2$, thus  for any arbitrary close solution (\ref{so1}), there are always continuously  closer solutions with smaller $\delta\varphi$ and $\delta\theta$ and hence   smaller $\sum_{i=1}^{3}C_{is}(\delta{\cal K}_{i})^2$,  which leads to smaller $\delta E$. Therefore,  none of the close solutions (\ref{so1}),  are  energetically stable. Note that, the close solutions (\ref{so1}) are those for which  the approximations (\ref{so3})  and (\ref{so4}) are valid.

\begin{figure}[ht!]
   \centering
   \includegraphics[width=110mm]{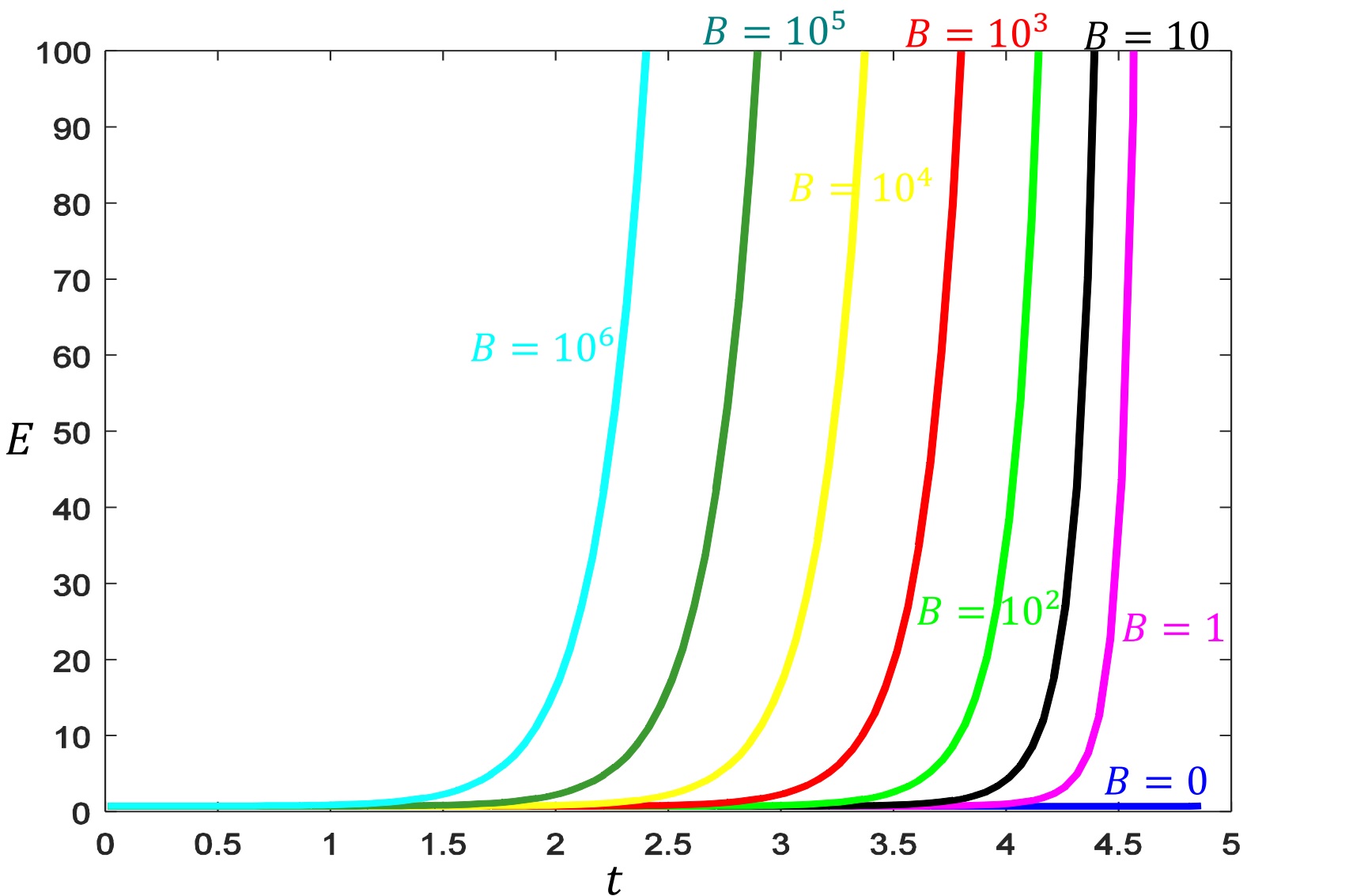}
   \caption{The variation  of the total energy $E$ versus time or all continuous  profiles which all together form    Fig.~\ref{pp}. Some of these  profiles are shown in   Fig.~\ref{3}. As  expected,  a horizontal line is obtained for  case $B=0$,  that is, the line $E=\pi/4$. } \label{2}
\end{figure}

Numerically, we should  study the stability of the SSWS (\ref{SSWS})  for some    arbitrary small hypothetical deformations.
For example, six  arbitrary slightly deformed  SSWSs  (\ref{so1}) can be introduced as follows:
\begin{equation}\label{var1}
\varphi(x,t)=\varphi_{s}+\delta \varphi=\frac{\pm 1}{\sqrt{1+ x^2}}+\xi~ \exp{(-x^{2})},\quad\quad \theta(x,t)=\omega_{s}t,
\end{equation}
 \begin{equation}\label{var2}
\varphi(x,t)=\varphi_{s}+\delta \varphi=\frac{\pm 1+\xi}{\sqrt{1+ x^2}},\quad\quad \theta(x,t)=\omega_{s}t,
 \end{equation}
\begin{equation}\label{var3}
\varphi(x,t)=\varphi_{s}+\delta \varphi=\frac{\pm 1}{\sqrt{ 1+\xi+ x^2}},\quad\quad \theta(x,t)=\omega_{s}t,
\end{equation}
 \begin{equation}\label{var4}
\varphi(x,t)=\varphi_{s}+\delta \varphi=\frac{\pm 1}{\sqrt{ 1+(1+\xi) x^2}},\quad\quad \theta(x,t)=\omega_{s}t,
\end{equation}
\begin{equation}\label{var5}
\varphi(x,t)=\frac{\pm 1}{\sqrt{1+ x^2}},\quad\quad \theta(x,t)=\theta_{s}+\delta \theta=\omega_{s}t+\xi~ t,
\end{equation}
\begin{equation}\label{var6}
\varphi(x,t)=\frac{\pm 1}{\sqrt{1+ x^2}},\quad\quad \theta(x,t)=\theta_{s}+\delta \theta=\omega_{s}t+\xi~ \exp{(-x^{2})},
\end{equation}
in which $\xi$ is a small parameter, which    can be an  indication  of  the order  of   deformations (variations) for any  kind  of  the small deformations (\ref{var1})-(\ref{var6}). All of the deformed functions (\ref{var1})-(\ref{var6})  turn to the same free non-deformed SSWS (\ref{SSWS}) for  $\xi=0$.  For all arbitrary deformations (\ref{var1})-(\ref{var6}),   Fig.~\ref{mmmm}(a-f) show  how  larger values of parameter  $B$ lead to more   stability. In other words, the larger  is the value of $B$ the greater would be the   increase in the total energy versus $|\xi|$. Figure~\ref{mmmm}(a-c) show  clearly  why the case $B=0$ leads  to an energetically  unstable SSWS. In other words, for the case $B=0$, in Figure.~\ref{mmmm}(a-c), the rest energy of the SSWS, i.e. $E_{s}=E(\xi=0)=\frac{\pi}{4}$, is not a minimum.   Note that, the case $B=0$ is  the same original RNKG system (\ref{lag}) with the same SSWS (\ref{ss}).

Some   hypothetical  deformations for the SSWS (\ref{SSWS})   can be considered  as those that appear in Figs.~\ref{pp} and \ref{3}. According to Fig.~\ref{pp}, the profile    of the initial SSWS (\ref{ss}) at $t=0$, does not  remarkably change at the
time interval $0 < t < 3$. In fact, it changes so slightly  that is not visible in Fig.~\ref{pp}. If one numerically  calculates
the total energy of the deformed SSWS   at different times
$0 < t < 5$,  Fig.~(\ref{2}) is obtained.  It reaffirms that for arbitrary small deformations above the background of the
SSWS (\ref{SSWS}), for example in the range $0 < t < 3$, the larger values of $B$ lead to more stability, that is, the larger the values  the greater will be the increase in the total energy for any arbitrary small variation above the background of the SSWS (\ref{SSWS}).
In general,    the total energy always increases (and  increases more for larger values of $B$) versus the amount of any arbitrary variation above the background of the SSWS (\ref{SSWS}). Although the parameter $B$ can be taken a  large value,  it would not affect the dominant dynamical equation (\ref{eom}) and the observable of the SSWS (\ref{SSWS}).

At the initial times (the times that are close to  $t=0$), a  multi lump (particle-like) solution with different velocities can be easily constructed just by adding distinct  far apart SSWSs (\ref{SSWS2}) together. For example, for $N$ distinct   SSWSs (\ref{SSWS2}) at different velocities $v_{j}$ and different initial positions $a_{j}$,  provided that $a_{j+1}-a_{j}\gg 1$, the multi particle-like solution at initial times  is as follows:
 \begin{equation}\label{SSWS3}
\varphi(x,t)=\sum_{j=1}^{N}\left[\frac{\pm 1}{\sqrt{1+\gamma_{j}^2(x-v_{j}t-a_{j})^2}}\right],
\end{equation}
where $\gamma_{j}=1/\sqrt{1-v_{j}^2}$. Since the phase field $\theta$ for each SSWS (\ref{SSWS2}) depends on its velocity, hence it must change from one to another.   That is to say,   if there are two SSWSs (i.e. $N=2$) with  one of them being at rest ($v_{1}=0$)  and the other  moving ($v_{2}\neq 0$), then the phase field must change from $\theta=\omega_{s}t$ at the position of the first SSWS  to  $\theta=k_{\mu}x^{\mu}=\omega_{2} t-k_{2}x$ at the position of the second SSWS. In the regions between two SSWSs, the scalar field $\varphi$ is almost zero and hence   $\varepsilon$ is almost zero everywhere.  Thus, there is not any rigorous restriction on $\theta$ to be in  the standard forms  $\theta=\omega_{s}t$ and $\theta=k_{\mu}x^{\mu}$ as  the special solutions of the PDE $\mathbb{S}_{1}=0$. In other words,  where the scalar  field   $\varphi$ is almost zero, the phase field $\theta$ is completely free and evolve without any rigorous restriction, i.e. it can change  slowly from $\theta=\omega_{s}t$ to $\theta=k_{\mu}x^{\mu}$  in  the spaces  for which  $\varphi\thickapprox 0$. In fact, for the case $\varphi\thickapprox 0$, it is not necessary    to satisfy   $\mathbb{S}_{i}=0$ or ${\cal K}_{i}=0$ ($i=1,2,3$) simultaneously, because for such situations, all $\varepsilon_{i}$'s ($i=1,2,3$) would automatically be almost zero simultaneously  without any restrictive condition.

The  general dynamical equations (\ref{geq}) and (\ref{geq2}), as two coupled nonlinear PDEs, have infinite solutions. Similar to the  SSWS (\ref{SSWS}), some of these solutions may be stable or there may not have any other stable solution at all. However,
if one considers  a system with a large  parameter $B$, as we have already shown, there is not any energetically stable solution among  the close solutions (\ref{so1}) at all.
But how is it possible  to know if the system has other stable solutions or not? In the present situation, it is by no means our goal to answer this question that can be mathematically important. In general, if one demonstrates that the other possible  stable solutions have very large energies, then physically, it is not an important issue, because they require  very large external energies  to be created and  it is outside the scope of our research. In fact,   except for  the  solutions which are very close\footnote{In order for a better understanding of the matter, we can provide a qualitative definition of  a conventional criterion.  We could designate  “\textit{very close solutions}" for those “\textit{close solutions}" (\ref{so1}) whose energy, for example for the case $B=10^{20}$, is less than $10^{3}$, or the ones for which  the magnitude of the variations $\delta\varphi$ and $\delta\theta$ is approximately less than $10^{-9}$.}  to the free far apart SSWSs (\ref{SSWS3}) and trivial vacuum state $\varphi=0$, the other (possible stable) solutions  can not be physically simply created. To put it differently, for any (possible stable) solution, it is not possible for all ${\cal K}_{i}$'s to be zero simultaneously, thus  at least one of  the $\varepsilon_{i}$'s, which contain the  large  parameter  $B$,  is a non-zero large function and then  the   energy is much  larger than the rest energy of a SSWS (\ref{SSWS}). Accordingly, the other (possible stable) solutions  need large  external  energies to be created, which is very unlikely to occur  physically. For example, if $B=10^{20}$, for a hypothetical small deformation (\ref{var2}) with  $\xi=\pm 10^{-5}$, the total energy is in the order of $10^{11}$, that is, the required  external energy must be of the order  $10^{11}$ to create  a small deformed SSWS (\ref{var2}) with  $\xi=\pm 10^{-5}$!

Furthermore, for other (possible stable) solutions, the additional  term $F$ is no longer zero  and is a large functional.  Thus, in the coupled PDEs (\ref{geq}) and (\ref{geq2}), the terms $\Box \varphi$ and $\frac{1}{2}\frac{dU}{d\varphi}$ are very small  compared with  the terms which contain $F$. In other words, the general   dynamical equations (\ref{geq}) and (\ref{geq2}), for the other solutions, which are not very close to the free far apart SSWSs (\ref{SSWS3}) and the vacuum state $\varphi=0$, are reduced to
\begin{eqnarray} \label{dfg}
&& \frac{\partial}{\partial x^{\mu}}\left(\frac{\partial F }{\partial (\partial_{\mu}\varphi)}\right)-\left(\frac{\partial F}{\partial \varphi}\right)  =0,\\ \label{dfg2} &&   \frac{\partial}{\partial x^{\mu}}\left(\frac{\partial F }{\partial (\partial_{\mu}\theta)}\right)=0,
\end{eqnarray}
as their dominant dynamical equations. In terms of functional ${\cal K}_{i}$'s, since $F=B\sum_{i=1}^{3}{\cal K}_{i}^{3}$, equivalently Eqs.~(\ref{dfg}) and (\ref{dfg2}),  turn into
\begin{eqnarray} \label{dlo}
&&\sum_{i=1}^{3} \left[2{\cal K}_{i}(\partial_{\mu}{\cal K}_{i})   \frac{\partial{\cal K}_{i}}{\partial(\partial_{\mu}\varphi)}    +   {\cal K}_{i}^2\partial_{\mu}\left(\frac{\partial{\cal K}_{i}}{\partial(\partial_{\mu}\varphi)}\right)    -    {\cal K}_{i}^2\frac{\partial{\cal K}_{i}}{\partial \varphi}   \right]=0,~~~~~~~\\&&\label{dlo1}
\sum_{i=1}^{3} \left[2{\cal K}_{i}(\partial_{\mu}{\cal K}_{i})   \frac{\partial{\cal K}_{i}}{\partial(\partial_{\mu}\theta)}    +   {\cal K}_{i}^2\partial_{\mu}\left(\frac{\partial{\cal K}_{i}}{\partial(\partial_{\mu}\theta)}\right)       \right]=0.
\end{eqnarray}
The important  point is that the parameter $B$ is again  ineffective in these equations. In other words, any solution of the coupled PDEs (\ref{dlo}) and (\ref{dlo1}) would be approximately  a solution of the general dynamical equations (\ref{geq}) and (\ref{geq2}) as well, provided that $B$ is considered to be a large number.

For the  far apart free SSWSs (\ref{SSWS3}), the dominant dynamical equation is the same simple equation (\ref{eom}),  and the role of the catalyzer term (\ref{FFF}) is approximately zero. However, when they get close to each other and
their profiles change slightly, the role of the stability catalyzer term  $F$ becomes important and strongly opposes a closer approach and causes  more change in their  profiles.
In this situation, the dominant dynamical equations are the same general forms  (\ref{geq}) and (\ref{geq2}).
Accordingly, as  expected  in the collision between the SSWSs (\ref{SSWS3}), they reappear  after collision processes without any significant  distortion.
For example,  for two SSWSs (\ref{SSWS3}) which  are initialized to collide with each other at the same speed $v=1-10^{-11}$, that is very close to the speed of light, the  kinetic energy of each one  is $K=E-E_{s}\thickapprox 2.2\times 10^{5}$. Now, if $B=10^{20}$, for a hypothetical deformation like (\ref{var2}), such  kinetic energy can only cause a  variation in the order of  $\xi\approx10^{-8}$ for each SSWS, that is, they reappear after the collision without any significant  distortion.  In general, for two free SSWSs (\ref{SSWS3}), while they are far apart, the total energy is finite, but when they come close together to interact, their profiles can   not have  notable  deformations, because their  initial kinetic energy must be a  huge  value  to  generate a  remarkable deformation for each SSWS, which is not  physically simple to be provided.

Based on all that has been said so far,  it is obvious that  if one considers  a system with an extremely   large value of $B$ (for example $B=10^{20}$ or even more),  the other  configurations of the fields $\varphi$ and $\theta$,   which are not very  close to the free far apart SSWSs (\ref{SSWS3}) and the vacuum state $\varphi=0$, require  extreme  energy to be created.  From a  physical  point of view, this issue can be interesting, as it classically explains how a system  leads to many identical particles with  specific characteristics. In other words, if one considers this system as a real physical system,   it is not possible to provide  an extremely large   external energy at a special place  for creating the other (possible stable) configurations of the fields. Thus, the only non-trivial stable configurations  of the fields  with the finite energies would be  any number of the  far apart SSWSs (\ref{SSWS3}) as a multi particle-like solution.
Similar to the  quantum field theory, the free far apart SSWSs (\ref{SSWS3})   can be  classically called the  quanta of the  system. Again, it should be noted  that, it does not matter to us whether the system has other possible  stable solutions.
What is important to us here is that the only non-trivial stable solutions with  finite energies are any number of the free far apart SSWSs (\ref{SSWS3}), as many identical particles with the specific characteristics,   which can be interesting for physicists.

\section{Summary and conclusion}\label{sec6}

In this paper, we introduced an extended  Klein-Gordon  system as an example (\ref{lag2}), which analytically yields an energetically stable solitary wave solution (\ref{SSWS}). In other words, it leads to a soliton solution. The new Lagrangian density (\ref{lag2}) is composed of two distinct parts, first, the original  part which is a known standard RNKG system (\ref{lag}), and second, an additional part, which can be called  the stability catalyzer term (\ref{FFF}). The original standard RNKG  Lagrangian density (\ref{lag}) is introduced for a single scalar field $\varphi$. But,  to introduce a proper   stability catalyzer term, it is necessary to use a different scalar field  $\theta$ (phase field) along with the original scalar field  $\varphi$, meaning that the stability catalyzer term is a functional of  $\varphi$ and $\theta$ simultaneously. The role of the  stability catalyzer term  seems as a massless spook   which surrounds the SSWS (\ref{SSWS}) and   guarantees the  stability of the SSWS (\ref{SSWS}). The stability catalyzer term  has no role in    the dominant   dynamical equation  and the other properties  of the SSWS (\ref{SSWS}), meaning that the general  dynamical equations (\ref{geq}) are reduced to the same   known standard RNKG version (\ref{eom}) just for the SSWS (\ref{SSWS}).  However, it guarantees   the energetical stability of the SSWS (\ref{SSWS}). Therefore,
any arbitrary small deformation above the background of  the SSWS (\ref{SSWS}) leads to an increase in the total energy. In other words, the rest energy of the SSWS is at a  minimum among the other solutions of the new extended KG system (\ref{lag2}) except for the ones which are very close to the vacuum state $\varphi=0$.

There is a parameter $B$ in the stability catalyzer term, the larger values of which, leads to more stability; meaning that, the larger the values the greater will be the increase in the total energy for any arbitrary small variation above the background of the SSWS (\ref{SSWS}). Hence,  considering a system with an extremely large value of $B$, leads to a classical  system  only with multi particle-like solutions. In fact,  the other   solutions of the system, which are not very close to the free far apart  SSWSs (\ref{SSWS3})  and the vacuum state $\varphi=0$, require infinite   amounts of energy to be created, and  are not physically possible to occur.  Thus, physically the possible stable solutions of the system with the finite energies, are either  any number of the far apart SSWSs (as any number of identical free particles) or the trivial  vacuum state. In other words, the SSWS (\ref{SSWS}) can be considered as the quantum of this classical system.

\section*{Acknowledgement}

The author  wishes to express his appreciation to the Persian Gulf University Research Council for their constant support.

\appendix

\section{}

Here, we are going to show that the following three PDE's
\begin{eqnarray} \label{true1}
  &&\mathbb{S}_{1}=\dot{\theta}^2-\theta'^2-1=0,\\\label{true2}&&
 \mathbb{S}_{2}=\dot{\varphi}^2-\varphi'^2+\varphi^{4}(1-\varphi^{2})=0,\\\label{true3}&&
 \mathbb{S}_{3}=\dot{\varphi}\dot{\theta}-\varphi'\theta'=0.
\end{eqnarray}
do not have any non-trivial common solutions except for the SSWS (\ref{SSWS}). Equation (\ref{true3}) generates $\dot{\theta}$ in terms of $\theta'$, $\varphi'$ and $\dot{\varphi}$ as follows:
\begin{equation}\label{fgh}
\dot{\theta}=\dfrac{\varphi'\theta'}{\dot{\varphi}}.
\end{equation}
If we insert this into Eq.~(\ref{true1}), we can obtain $\theta'$ in terms of $\varphi'$ and $\dot{\varphi}$ as follows:
\begin{equation}\label{fgh2}
\theta'=\dfrac{\pm\dot{\varphi}}{\sqrt{\varphi'^2-\dot{\varphi}^2}}.
\end{equation}
Using Eqs. (\ref{fgh}) and (\ref{fgh2}), $\dot{\theta}$ can be obtained  as well:
\begin{equation}\label{fgh3}
\dot{\theta}=\dfrac{\pm\varphi'}{\sqrt{\varphi'^2-\dot{\varphi}^2}}.
\end{equation}
The obvious mathematical  expectation $(\dot{\theta})'=\dfrac{d}{dx}\dfrac{d\theta}{dt}=\dfrac{d}{dt}\dfrac{d\theta}{dx}=\dot{(\theta')}$
leads to the following result:
\begin{equation}\label{vgh}
\ddot{\varphi}-\varphi''+\dfrac{1}{\sqrt{\varphi'^2-\dot{\varphi}^2}}
(\dot{\varphi}^2\ddot{\varphi}+\varphi'^2\varphi''-2\dot{\varphi}\varphi'\dot{\varphi}')=0,
\end{equation}
which can be simply written  in a covariant form:
\begin{equation}\label{fjh}
\partial_{\mu}\partial^{\mu}\varphi+\frac{1}{\sqrt{-\partial_{\mu}\varphi\partial^{\mu}\varphi}}(\partial_{\nu}\varphi\partial_{\sigma}\varphi)
(\partial^{\nu}\partial^{\sigma}\varphi)=0
\end{equation}
Therefore, to find  the common solutions of the three independent nonlinear PDE's (\ref{true1}), (\ref{true2}) and (\ref{true3}), equivalently  we can  search  for the common solutions of the    two different PDE's (\ref{true2}) and (\ref{fjh}).  In general, it is easy to show that each non-vibrational   function  $\varphi_{v}(x,t)=\varphi_{o}(\gamma(x-vt))$, would be a solution of the  PDE (\ref{fjh}) or (\ref{vgh}). Moreover, for any non-vibrational solitary wave solution, Eqs. (\ref{fgh2}) and (\ref{fgh3}) lead to $\theta'=\pm \gamma v=\omega_{s}\gamma v=\omega v$ and $\dot{\theta}=\pm\gamma =\gamma\omega_{s}=\omega$ as we expected.
On the other hand, we know that the SSWS (\ref{SSWS2}) is the single  non-vibrational  solution of the PDE (\ref{true2}). Hence, for PDE's (\ref{true2}) and (\ref{fjh}), the single common non-vibrational  solitary wave solution is the SSWS (\ref{SSWS2}), as we expected. Accordingly, for the scalar  field $\varphi$, there are two completely different   PDE's (\ref{true2}) and (\ref{fjh}). Therefore, it does  not seem that  other common vibrational solutions exist along with the single non-vibrational SSWS (\ref{SSWS2}).

\end{document}